\documentstyle[12pt,aasms]{article}

\begin{document}

\def\LSUN{\rm L_{\odot}}
\def\MSUN{\rm M_{\odot}}
\def\RSUN{\rm R_{\odot}} 
\def\MSUNYR{\rm M_{\odot}\,yr^{-1}}
\def\MSUNS{\rm M_{\odot}\,s^{-1}}
\def\MDOT{\dot{M}}

\newbox\grsign \setbox\grsign=\hbox{$>$} \newdimen\grdimen \grdimen=\ht\grsign
\newbox\simlessbox \newbox\simgreatbox
\setbox\simgreatbox=\hbox{\raise.5ex\hbox{$>$}\llap
     {\lower.5ex\hbox{$\sim$}}}\ht1=\grdimen\dp1=0pt
\setbox\simlessbox=\hbox{\raise.5ex\hbox{$<$}\llap
     {\lower.5ex\hbox{$\sim$}}}\ht2=\grdimen\dp2=0pt
\def\simgreat{\mathrel{\copy\simgreatbox}}
\def\simless{\mathrel{\copy\simlessbox}}

\title{Axisymmetric MHD simulations of the collapsar model
for gamma-ray bursts}

\vspace{1.cm}
\author{Daniel Proga$^{1}$, Andrew I. MacFadyen$^{2}$, Philip J. Armitage$^{1, 3}$, 
and Mitchell C. Begelman$^{1, 3}$} 
\vspace{.5cm}
\affil{$^1$JILA, University of Colorado, Boulder, CO 80309-0440, USA}
\affil{$^2$California Institute of Technology, Mail Code 130-33, 
Pasadena, CA 91125}
\affil{$^3$also Department of Astrophysical and Planetary Sciences, University
of Colorado, Boulder}

\vspace{2cm}

\centerline{submitted to ApJ Letters on Aug. 28 2003}

\begin{abstract}
We present results from axisymmetric, time-dependent
magnetohydrodynamic (MHD) simulations of the collapsar model for
gamma-ray bursts.  We begin the simulations after the $1.7~\MSUN$ iron
core of a 25~$\MSUN$ presupernova star has collapsed and study the
ensuing accretion of the $7~\MSUN$ helium envelope onto the central
black hole formed by the collapsed iron core.  We consider a
spherically symmetric progenitor model, but with spherical symmetry
broken by the introduction of a small, latitude-dependent angular
momentum and a weak radial magnetic field. Our MHD simulations include
a realistic equation of state, neutrino cooling, photodisintegration
of helium, and resistive heating.  Our main conclusion is that, within
the collapsar model, MHD effects alone are able to launch, accelerate
and sustain a strong polar outflow.  We also find that the outflow is
Poynting flux-dominated, and note that this provides favorable initial
conditions for the subsequent production of a baryon-poor fireball.

\end{abstract}


\section{Introduction}
The collapsar model is one of most promising scenarios to explain the
huge release of energy in a matter of seconds, associated with
gamma-ray bursts (GRBs; Woosley 1993; Paczy\'{n}ski, 1998; MacFadyen
\& Woosley 1999, hereafter MW; Popham, Woosley \& Fryer 1999;
MacFadyen, Woosley \& Heger 2001). In this model, the collapsed iron
core of a massive star accretes gas at a high rate ($\sim 1 \MSUNS$)
producing a large neutrino flux, a powerful outflow, and a GRB.
Despite many years of intensive theoretical studies of these events,
basic properties of the central engine are uncertain. In part, this is
because previous numerical studies of the collapsar model did not
explicitly include magnetic fields, although they are commonly
accepted as a key element of accretion flows and outflows.

In this letter we present a study of the time evolution of
magnetohydrodynamic (MHD) flows in the collapsar model. This study is
an extension of existing models of MHD accretion flows onto a black
hole (BH; Proga \& Begelman 2003, PB03 hereafter).  In particular, we
include a realistic equation of state (EOS), photodisintegration of
bound nuclei and cooling due to neutrino emission.  Our study is also
an extension of MW's collapsar simulations, as we consider very similar
neutrino physics and initial conditions but solve MHD instead of
hydrodynamical equations.

\section{Method}

To calculate the structure and evolution of an accreting flow, we solve 
the equations of MHD:
\begin{equation}
   \frac{D\rho}{Dt} + \rho \nabla \cdot {\bf v} = 0,
\end{equation}
\begin{equation}
   \rho \frac{D{\bf v}}{Dt} = - \nabla P - \rho \nabla \Phi+ \frac{1}{4\pi} {\bf (\nabla \times B) \times B},
\end{equation}
\begin{equation}
   \rho \frac{D}{Dt}\left(\frac{e}{ \rho}\right) = -P \nabla \cdot {\bf
   v}+\eta_r\bf{J}^2 -\cal{L},
\end{equation}
\begin{equation}
{\partial{\bf B}\over\partial t} = {\bf\nabla\times}({\bf v\times B-
\eta_{\it r} J}),
\end{equation}
where $\rho$ is the mass density, $P$ is the total gas pressure plus radiation
pressure, 
${\bf v}$ is the fluid velocity, $e$ is the internal energy density,
$\Phi$ is the gravitational potential, $\bf B$ is the magnetic field vector,
$\bf J$ is the current density, $\eta_r$ is an anomalous resistivity,
and $\cal{L}$ is the cooling rate due to neutrinos.

To compute resistivity, we follow Stone \& Pringle (2001, see their
equations 5 and A1).  We perform simulations using the
pseudo-Newtonian potential of the central mass $\Phi_{pw} =
GM/(r-R_S)$, where $R_S = 2GM/c^2$ is the Schwarzschild radius,
introduced by Paczy\'{n}ski \& Wiita (1980).  We increase the mass of
the BH during the calculation by the amount of 
baryonic rest mass 
accreted through the inner boundary.  

Our calculations are performed in spherical polar coordinates
$(r,\theta,\phi)$. We assume axial symmetry about the rotational axis
of the accretion flow ($\theta=0^\circ$ and $180^\circ$).  The
computational domain is defined to occupy the radial range
$r_i~=~1.5~R_S \leq r \leq \ r_o~=~ 1000~R_S$, and the angular range
$0^\circ \leq \theta \leq 180^\circ$. The $r-\theta$ domain is
discretized on a non-uniform grid as in PB03, which yields $\Delta
r/r=0.278$ at the inner edge of the simulations.

We adopt a realistic EOS, which includes contributions from an ideal
gas of nuclei, radiation, and electrons and positrons with arbitrary
degrees of relativity and degeneracy (Blinnikov, Dunina-Barkovskaya \&
Nadyozhin 1996).  To compute the cooling rate, we follow Itoh et
al. (1989, 1990), taking into account thermal neutrino emission
processes dominated by pair annihilation, as well as the capture of
pairs on free nucleons.

Our calculations use the ZEUS-2D code described by Stone \& Norman
(1992a,b).  We have extended the code to include the realistic EOS,
artificial resistivity, photodisintegration and neutrino cooling. We
included the terms due to the resistivity and cooling in an operator
split fashion separately from the rest of the dynamical equations. For
stability, these terms must be integrated using the time step computed
based on the resistive and cooling time scales, respectively. We
subcycle whenever either of these two time steps is smaller than the
time step used in the MHD equations (see, e.g., Stone, Pringle, and
Begelman 1999).  Inclusion of a non-adiabatic EOS requires iterating over
temperature, $T$ (see MW for details).

By including the neutrino cooling and realistic EOS, we consider very
similar microphysics to that used in MW.  Our simulations differ from
those in MW in that we use the ZEUS MHD code whereas MW used the
PROMETHEUS hydrodynamics code (Fryxell, M$\ddot{\rm u}$ller \& Arnett
1989).  This means that we can self-consistently calculate turbulent
stresses generated by the magnetorotational instability (MRI) and thus
include the outward transport of energy and angular momentum (e.g.,
Balbus \& Hawley 1991).  MW implemented the
effects of viscosity in the disk using an alpha viscosity as
prescribed by Shakura \& Sunyaev (1973).  We do not consider
self-gravity and nuclear burning; however, as in MW our simulations
track the photodisintegration of helium.  Our simulations span the radial
direction from 1.5 to 1000 $R_S$ whereas MW's simulations span from
9.5 to 9500 $R_S$.  Thus we can capture the innermost part of the flow
near the BH but follow the evolution out to a smaller radius than MW's
simulations.

We adopt PB03's boundary conditions and initial conditions for the
magnetic field. In particular, the initial magnetic field is purely
radial and weak ($\beta \equiv 8\pi P/ B^2 \gg 1$ everywhere).  For
the initial conditions of the fluid variables, we follow MW and adopt
the stellar model for the helium core of a 25~$\MSUN$ presupernova
star (model S251S7B@14233 in Woosley \& Weaver 1995).  The masses of
the helium and iron core derived from this star are 7.23 and
1.70~$\MSUN$, respectively.  Similarly to MW's simulations, our
simulations start after the entire iron core is assumed to have
collapsed to form a BH (with $R_S=4.957\times 10^5$~cm), but before
the helium envelope has collapsed. The model predicts an inner radius
of the helium envelope, $R_{\rm He}$, of 2.11$\times 10^8$~cm. Outside
this radius, we adopt the radial velocity from the stellar model, set
$v_\theta=0$ and assume a non-zero $l$.  The angular momentum
distribution is chosen such that the ratio between the centrifugal
force and the component of gravity perpendicular to the rotational
axis is 0.02 at all angles and radii, except where this prescription
results in $l > l_{max}=10^{17}~{\rm cm^2~s^{-1}}$; then we set $l=
l_{max}$.  Inside $R_{\rm He}$, we set $v_\theta=v_\phi=0$ and assume
a free-fall radial velocity. We compute the initial density inside the
helium envelope using the mass continuity equation and the mass
accretion rate from the stellar model at $R_{\rm He}$, where
$\rho=1.16 \times 10^7~{\rm g~cm^{-3}}$ and $v_r=-8.81\times10^7~{\rm
cm~s^{-1}}$.

\section{Results}

With these assumptions, there is only one free parameter which
defines the strength of the initial magnetic field. In this letter,
we present results from a single model for which $\beta=10^6$
at the outer boundary.

We find that after a transient episode of infall, lasting 0.13 s, the
gas with $l\simgreat 2 R_S c$ piles up outside the black hole and
forms a thick torus bounded by a centrifugal barrier near the rotation
axis.  Soon after the torus forms (i.e., a couple of orbits at
$r=r_i$), the magnetic field is amplified by both MRI and shear.
We have verified that most of the inner torus is unstable to MRI, and
that our simulations have enough resolution to resolve, albeit
marginally, the fastest growing MRI mode.  The torus starts evolving
rapidly and accretes onto the black hole. Another important effect of
magnetic fields is that the torus produces a magnetized corona and an
outflow.  The presence of the corona and outflow is essential to the
evolution of the inner flow at all times and the entire flow close to
the rotational axis during the latter phase of the evolution.  We find
that the outflow very quickly becomes sufficiently strong to overcome
supersonically infalling gas (the radial Mach number in the polar
funnel near the inner radius is $\sim 5$) and makes its way outward,
reaching the outer boundary at $t=0.25$ s. Due to limited computing
time, our simulations were stopped at $t=0.28215$~s, which corresponds
to 6705 orbits of the flow near the inner boundary.  We
expect the accretion to continue much longer, roughly the collapse
timescale of the Helium core ($\sim 10$~s), as in MW.

Figure~1 shows the time evolution of the mass accretion rate through
the inner boundary (left panel), total magnetic energy (second left
panel), neutrino luminosity (third left panel) and radial Poynting and
kinetic flux along the polar axis at $r=190~R_S$ (right panel).
Unless otherwise stated, all quantities in this paper are in cgs
units.  Initially, during a precollapse phase, $\MDOT_a$ stays nearly
constant at the level of $\sim 5\times 10^{32}~{\rm g~s^{-1}}$. During
this phase the zero-$l$ gas inside the initial helium envelope is
accreted.  Around $t=0.13$~s, $\MDOT_a$ rises sharply as the gas from
the initial helium envelope reaches the inner boundary. However, this
gas has non-zero $l$ and a rotational supported torus and its corona
and outflow form, causing a drop in $\MDOT_a$ after it reaches a
maximum of $2\times 10^{33}~{\rm g~s^{-1}}$ at $t=0.145$~s. The
accretion rate reaches a minimum of $6\times10^{31}~{\rm g~s^{-1}}$ at
$t\approx0.182$~s and then fluctuates with a clear long-term increase.
This increase is caused by the contribution from gas with $l< 2 R_S
c$, which is directly accreted (without need to transport $l$) from
outside the main body of the torus. The total mass and angular
momentum accreted onto the BH during our simulation (0.3 s) are
$0.1~\MSUN$ and $3\times10^{39}~{\rm g~cm^2~s^{-1}}$, respectively.

The time evolution of the total magnetic energy (integrated over the
entire computational domain) is characteristic of weakly magnetized
rotating accretion flows. Most of the magnetic energy is due to the
toroidal component of the magnetic field.  We note a huge increase of
the toroidal magnetic field coinciding with the formation and
development of the torus. Both $B_\phi$ and $B_\theta$ are practically
zero during the precollapse phase of the evolution. But at $t=0.14$~s
the total energy in $B_\phi$ equals that in $B_r$ and just 0.025~s
later the $B_\phi$ energy is higher than the $B_r$ energy by a factor
of 50. At the end of simulations the total kinetic energy from the
radial, latitudinal and rotational motion are $4\times 10^{50}$,
$6.5\times 10^{49}$, and $2.3\times10^{51}$~erg, respectively.  These
gross properties indicate that the magnetic energy is large enough to
play an important role in the flow dynamics.

The time evolution of the neutrino luminosity, $L_\nu$, shows that the
neutrino emission stays at a relatively constant level of
$3\times10^{52}~{\rm erg~s^{-1}}$ after the torus forms.  We compute
$L_\nu$ under the assumption that all the gas in the model is
optically thin to neutrinos, so that $L_\nu$ is volume integrated
$\cal{L}$ over the entire computational domain. We note that $L_\nu$
is dominated by the neutrino emission due to pair capture on free
nucleons (the so-called URCA cooling).

The last panel in Fig.~1 shows the area-integrated radial
fluxes of  magnetic and kinetic energy at $r=190~R_S$ inside
the polar outflow. 
The outflow is Poynting flux-dominated, with the Poynting
flux exceeding the kinetic energy flux by up to an order of magnitude.

Our analysis of the inner flow shows that the outflow is magnetically
driven from the torus.  Soon after the torus forms, the magnetic field
very quickly deviates from its initial radial configuration due to MRI
and shear.  This leads to fast growth of the toroidal magnetic field
as field lines wind up due to the differential rotation.  As a result
the toroidal field dominates over the poloidal field and the gradient
of the former drives an outflow.  Figure~2 shows the flow pattern of
the inner part of the flow at $t=0.2735$~s.  The left and right panel
show density and $|B_\phi|$ maps, respectively.  The maps are overlaid
by the direction of the poloidal velocity. The polar regions of low
density and high $B_\phi$ coincide with the region of an outflow. We
note also that during the latter phase of the evolution not all of the
material in the outflow originated in the innermost part of the torus
-- a significant part of the outflow is ``peeled off'' the infalling
gas at large radii by the magnetic pressure.  Figure~2
illustrates that the inner torus and its corona and outflow cannot
always prevent the low-$l$ gas from reaching the BH.  Even the
magnetic field cannot do it if the density of the incoming gas is too
high. Therefore, we find that the outflow of the magnetic energy
(mostly toroidal) from the innermost part of the flow does not always
correspond to an outflow of gas (in other words, the Poynting flux
and kinetic energy flux can be in opposite directions).

Figure~3 shows the radial profiles 
of several quantities in our run, angle-averaged over a small wedge near
the equator (between $\theta=86^\circ$ and $94^\circ$), and time-averaged
over 50 data files covering a period at the end of the simulations
(from 0.2629~s  through 0.2818~s).
We indicate the location of the last
stable circular orbit by the vertical dotted line in each panel.

The profiles of each variable are not simple power-laws but are rather
complex. In particular, the density has a prominent maximum of
$4\times10^{11}~{\rm g~cm^{-3}}$ at $r=5 R_S$.  The gas plus radiation
pressure is higher than the magnetic pressure. The rotational velocity
is nearly Keplerian inside $r=6 R_S$ and sub-Keplerian outside this
radius.

We measure the Reynolds stress, 
$\alpha_{gas}\equiv<\rho v_r \delta v_\phi>/P$, 
and the Maxwell stress normalized to the magnetic pressure, 
$\alpha_{mag}\equiv <2B_rB_\phi/B^2>$. 
Note that Figure~3 shows only the magnitude, 
not the sign, of the normalized stresses.
We find that except for $r\simless 2.5 R_S$ and 
$10 R_S\simless r \simless 12 R_S$,
the Maxwell stress dominates over the Reynolds
stress in the inner flow.
The last panel in Figure~3 shows that the toroidal component
of the magnetic field is dominant for $r< 50~R_S$.

We have compared the cooling time scale and the advection
time scale in the flow and found that overall the flow
is advection-dominated expect for a  small region inside
the torus where the density reaches its maximum.

\section{Discussion and Conclusions}

We have performed time-dependent two-dimensional MHD simulations of
the collapsar model.  Our simulations show that: 1) soon after the
rotationally supported torus forms, the magnetic field very quickly
starts deviating from purely radial due to MRI and shear. This leads
to fast growth of the toroidal magnetic field as field lines wind up
due to the torus rotation; 2) The toroidal field dominates over the
poloidal field and the gradient of the former drives a polar outflow
against supersonically accreting gas through the polar funnel; 3) The
polar outflow is Poynting flux-dominated; 4) The polar outflow reaches
the outer boundary of the computational domain ($5\times10^8$~cm) with
an expansion velocity of 0.2 c; 6) The polar outflow is in a form of
a relatively narrow jet (when the jet breaks through the outer
boundary its half opening angle is $5^\circ$); 7) Most of the energy
released during the accretion is in neutrinos, $L_\nu=2\times
10^{52}~{\rm erg~s^{-1}}$. Therefore it is likely that neutrino
driving can increase the outflow energy (e.g., Fryer \&
M\'{e}sz\'{a}ros 2003 and references therein).

Our simulations explore a relatively conservative case where we allow
for neutrino emission but do not allow for the emitted neutrinos
to interact with the gas or annihilate.  The only sources of
nonadiabatic heating in our simulations are the artificial viscosity
and resistivity.

Our main conclusion is that, within the collapsar model, MHD effects
are able to launch, accelerate and sustain a strong polar outflow. We
believe that this conclusion will turn out to be largely independent
of the initial magnetic field strength in the stellar core, because
MRI can rapidly amplify weak fields until they are strong enough to
drive a powerful outflow. Since our simulations are non-relativistic,
and cover only the innermost region of the collapsing star, we cannot
determine whether our outflows are sufficient to yield a
GRB. Additional driving could also be necessary. We also find that the
outflow is Poynting flux-dominated, and note that this provides
favorable initial conditions for the subsequent production of a
baryon-poor fireball [e.g., Fuller, Pruet \& Abazajian (2000);
Beloborodov (2003); Vlahakis \& K$\ddot{\rm o}$nigl (2003);
M\'{e}sz\'{a}ros (2002)], or a magnetically dominated ``cold
fireball'' [Lyutikov \& Blandford (2002)].

ACKNOWLEDGMENTS: 
DP acknowledges support from NASA under LTSA grants NAG5-11736 
and NAG5-12867.
MCB acknowledges support from NSF grants 
AST-9876887 and AST-0307502.
\newpage
\section*{ REFERENCES}
 \everypar=
   {\hangafter=1 \hangindent=.5in}

{

  Balbus, S. A., \& Hawley, J. F. 1991, ApJ, 376, 214

 Beloborodov, A.M. 2003, ApJ, 588, 931

 Blinnikov, S.I., Dunina-Barkovskaya, N.V., \& Nadyozhin, D. K. 
 1996, ApJS, 106, 171

 Fuller, G. M., Pruet, J., \& Abazajian, K. 2000, Phys. Rev. Lett., 85, 2673

 Fryxell, B. A., M$\ddot{\rm u}$ller, E., \& Arnett, W. D. 1989, MPA Rep. 449 (Garching: MPA)

 Fryer, C. L., \& M\'{e}sz\'{a}ros, P. 2003, ApJ,  588, L25

 Itoh, N., Adachi, T., Nakagawa, M., Kohyama, Y., \&
 Munakata, H. 1989, ApJ, 339, 354

 Itoh, N., Adachi, T., Nakagawa, M., Kohyama, Y., \&
 Munakata, H. 1990, ApJ, 360, 741

 Lyutikov, M. \& Blandford, R. 2002, APS, APR, 6008

 MacFadyen, A., \& Woosley, S.E. 1999, ApJ, 524, 262  (MW)

 MacFadyen, A., Woosley, S.E., \& Heger A. 2001, ApJ, 550, 410

  M\'{e}sz\'{a}ros, P.  2002, ARA\&A, 40, 137

  Paczy\'{n}ski, B. 1998, ApJ, 494, L45

 Paczy\'{n}ski, B., \& Wiita, P. J. 1980, A\&A, 88, 23 (PW)

 Popham, R., Woosley, S.E., \& Fryer, C. 1999, ApJ, 518, 356 

 Proga, D., \& Begelman, M.C. 2003 ApJ, 592, 767 (PB03)

 Shakura, N.I., \& Sunyaev, R.A. 1973 A\&A, 24, 337

  Stone, J.M., \& Norman, M.L. 1992a, ApJS, 80, 753

  Stone, J.M., \& Norman, M.L. 1992b, ApJS, 80, 791

 Stone, J. M., \& Pringle, J. E. 2001, MNRAS, 322, 461

 Stone, J.M., Pringle, J.E., \& Begelman, M.C. 1999, MNRAS, 310, 1002

 Vlahakis, N., \& K$\ddot{\rm o}$nigl, A. 2003, ApJ, submitted

 Woosley, S.E. 1993, ApJ, 405, 273 

 Woosley, S. E., \& Weaver, T. A. 1995, ApJS, 101, 181

}

\eject

\newpage
\centerline{\bf Figure Captions}

\vskip 4ex
\noindent
Figure~1 -- The time evolution of the mass accretion rate (left panel),
total magnetic energy due to each of the three field components
(second left panel), neutrino luminosity 
(third left panel) 
and area-integrated radial Poynting and kinetic flux in the polar outflow at
$r=190~R_S$ (right panel). Formally, we define the polar
outflow as the region where $v_r>0$ and $\beta<1$.
Note the difference in the time range in the panel with the
radial fluxes.

The last panel in Fig.~1 shows the area-integrated radial
fluxes of  magnetic and kinetic energy at $r=190~R_S$ inside
the polar outflow. 
Formally, we define the polar
outflow as the region where $v_r>0$ and $\beta<1$.
The outflow is Poynting flux-dominated, with the Poynting
flux exceeding the kinetic energy flux by up to an order of magnitude.

\vskip 4ex
\noindent
Figure~2 -- 
Color maps of logarithmic density and toroidal magnetic field 
overplotted with the direction
of the poloidal velocity at $t=0.2735$~s.
The length scale is in units of the BH radius (i.e., $r'=r/R_S$
and $z'=z/R_S$).

\vskip 4ex
\noindent
Figure~3 -- 
Radial profiles of various quantities from our run, 
time-averaged 
from $0.2629$  through $0.2818$~s. 
To construct each plot, we averaged the profiles
over angle between $\theta=86^\circ$ and $94^\circ$.
The top left panel plots the density (solid line) and temperature (dashed line).
The top middle panel plots the gas pressure (solid line) and magnetic
pressure. The top right panel plots the rotational, radial,
Keplerian, and Alfv${\acute{\rm e}}$n velocities  
(solid, dashed, dot-dashed, and dotted line, respectively), as well as
the sound speed (triple-dot dashed line).
The bottom left panel plots the angular velocity in units of $2c/R_s$.
The bottom middle panel plots the Maxwell stress,  
$\alpha_{mag}$,
and the Reynolds stress, $\alpha_{gas}$
(solid and dashed line, respectively). 
We calculate the Reynolds stress using eq. (15) in PB03 
and  show only its amplitude.
The bottom right panel plots
the radial, latitudinal and toroidal components of the magnetic field
(dot-dashed, dashed, and solid line, respectively).
The length scale is in units of the BH radius (i.e., $r'=r/R_S$).

\eject
\newpage

\begin{figure}
\begin{picture}(180,590)

\put(280,423){\includegraphics{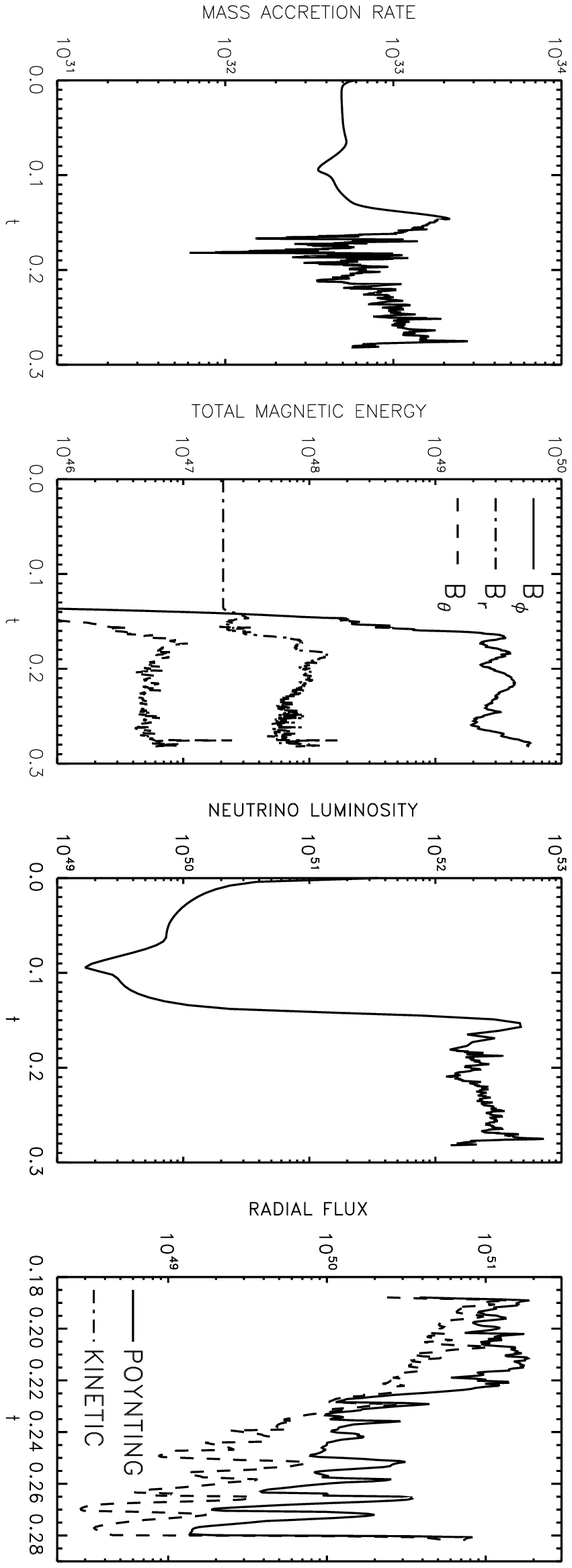}}

\end{picture}
\caption{ }
\end{figure}

\eject
\newpage

\begin{figure}
\begin{picture}(180,590)

\put(280,423){\includegraphics{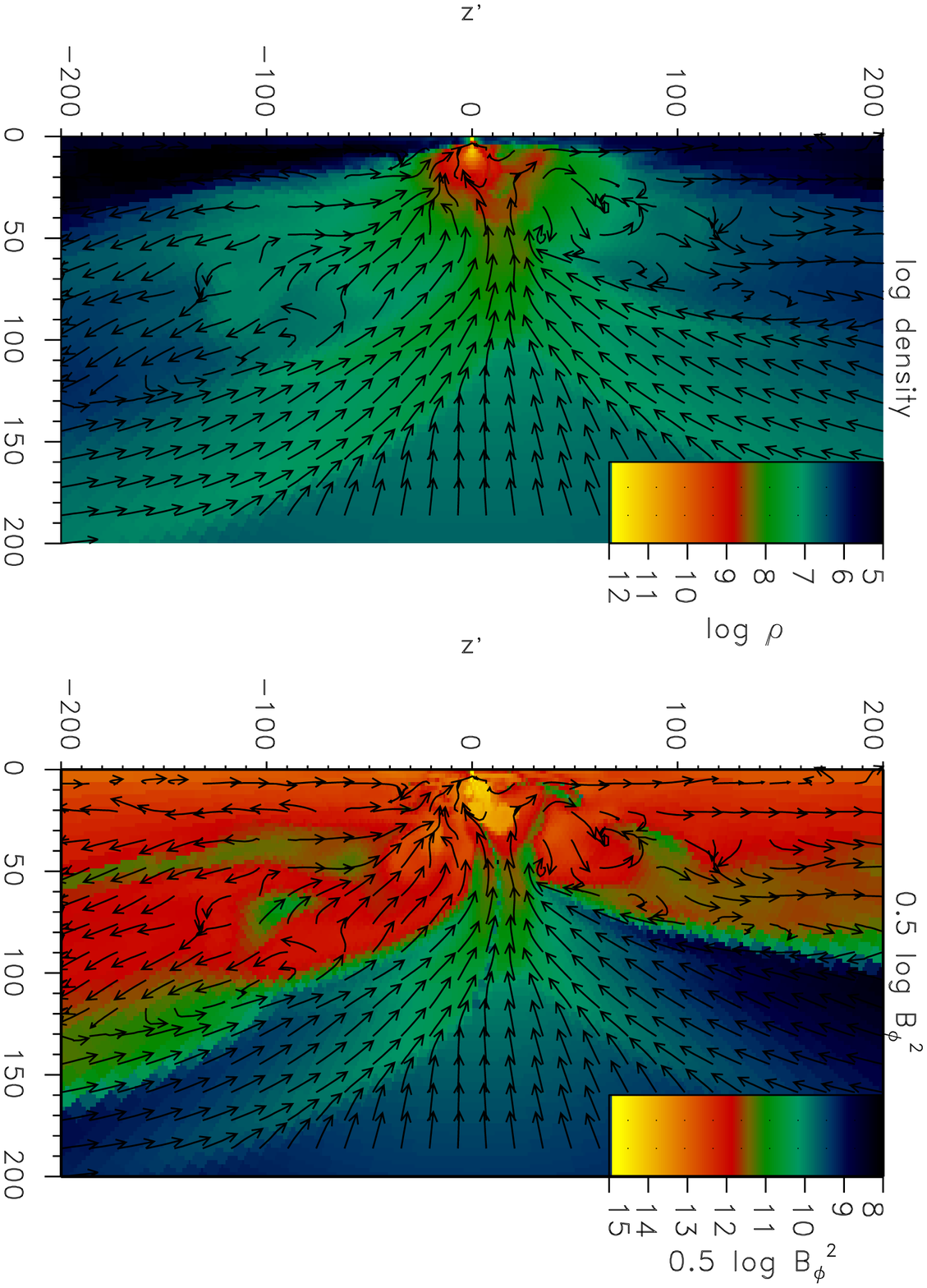}}

\end{picture}
\caption{}
\end{figure}

\eject
\newpage

\begin{figure}
\begin{picture}(180,590)

\put(280,423){\includegraphics{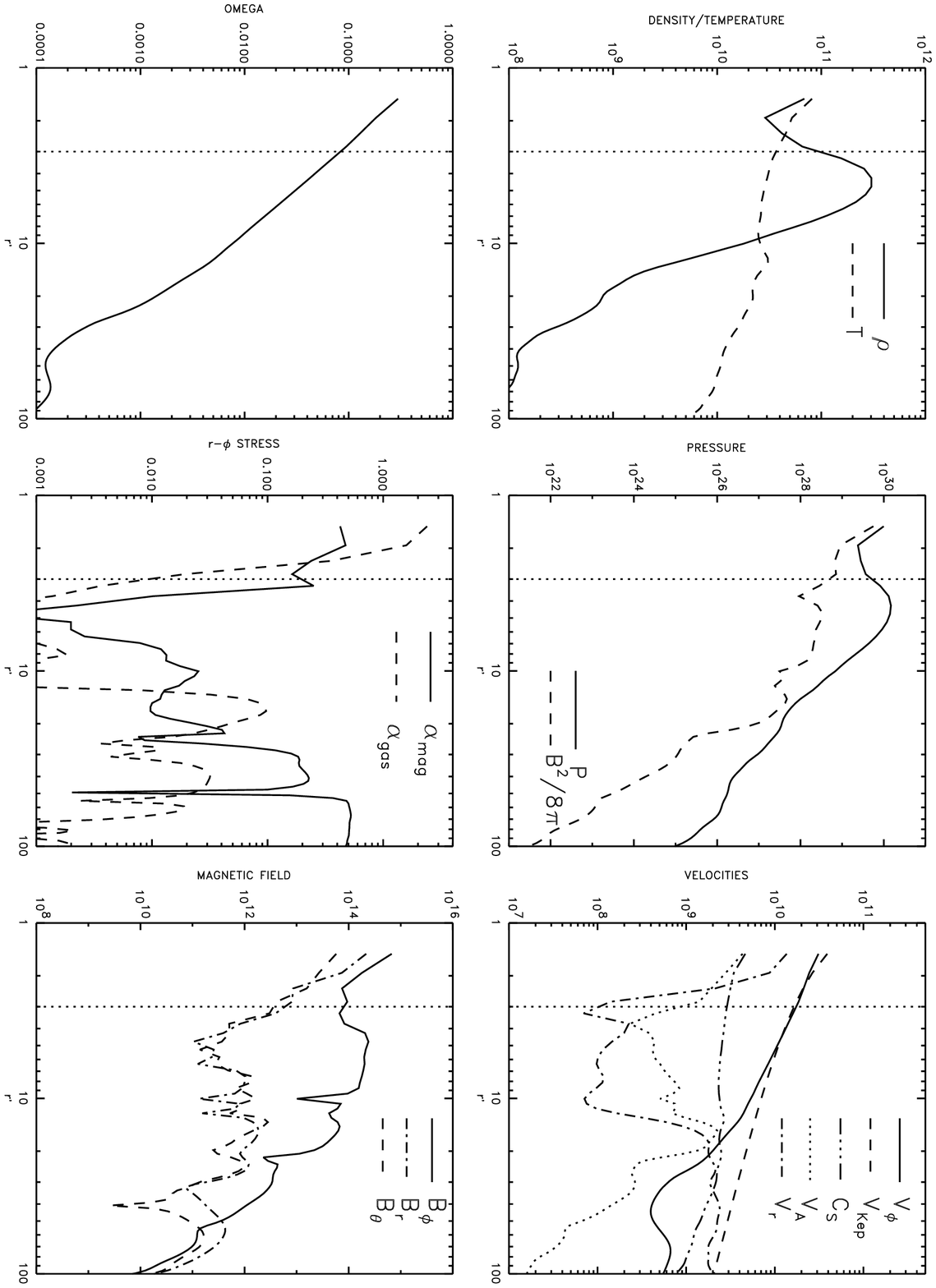}}

\end{picture}
\caption{}
\end{figure}

\end{document}